\documentclass[aps,apl,onecolumn,superscriptaddress,showpacs,11pt]{revtex4-1}
\usepackage{graphicx}
\usepackage{dcolumn}
\usepackage{bm}
\usepackage[latin1]{inputenc}
\usepackage{amsfonts}
\usepackage{amssymb}
\usepackage{amsmath}
\usepackage{color}
\usepackage{ulem}
\usepackage{multirow}
\usepackage{booktabs}
\begin{document}

\title{Magneto-optical Properties of Reduced Titania Probed by First-principles Calculations}     

\author{C. Echeverr\'ia-Arrondo}
\affiliation{Centro de F\'isica de Materiales, CSIC-CFM, San Sebasti\'an, Spain}
\author{H. Raebiger}
\affiliation{Physics Department, Yokohama University, Yokohama, Japan}
\author{J. P\'erez-Conde}
\affiliation{Dpto. de Ciencias. Universidad P\'ublica de Navarra. 31006 Pamplona. Spain}
\author{C. G\'omez-Polo}
\affiliation{Dpto. de Ciencias. Universidad P\'ublica de Navarra. 31006 Pamplona. Spain}
\affiliation{Institute for Advanced Materials (INAMAT), Universidad P\'ublica de Navarra, 31006 Pamplona, Spain}
\author{A. Ayuela}
\affiliation{Centro de F\'isica de Materiales, CSIC-CFM, San Sebasti\'an, Spain}

\email{cechevar@uji.es}

\begin{abstract}
The magneto-optical properties of titanium dioxide systems are related to the presence of impurity states in the band gap due to oxygen vacancies. To understand about the interplay between localized electrons and structural distortions at the vacancy sites and the magneto-optical properties, we employ a self-interaction corrected density functional theory method to calculate bulk and small nanoparticles of rutile, anatase, and brookite titania. Our computations reveal bipolaron configurations associated to an oxygen vacancy with optical transition levels in the band gap.  The ground state for these bipolarons is a spin-triplet state in bulk rutile TiO$_2$ and also in the nanoparticles independently of the crystal phase, a result which may support the idea of oxygen vacancies as a source of magnetism in this material. The ground state for bipolarons in bulk anatase TiO$_2$ is however a spin-singlet state, different from the spin-triplet configuration reported in a previous work based on hybrid functionals. 
\end{abstract}
\maketitle

\section{Introduction}
Titanium dioxide is a wide band gap semiconductor metal oxide of great technological relevance for applications including perovskite solar cells\cite{pvk1,pvk2,mio}, photocatalysis\cite{nanoscale,photo}, and spintronic devices\cite{spin1,spin2,spintronics}. Titanium dioxide doped with light elements such as N\cite{cristina,gp,doped1,expart2} and C\cite{expart1} and also with transition metal dopants shows ferromagnetism with Curie temperature above room temperature. Ferromagnetism is still observed in undoped crystals such as TiO$_2$ samples formed under strongly reducing synthesis conditions, wherein oxygen vacancies behave as intrinsic donors.\cite{cristina,defect1,defect3,expart3}

In this paper, to further understand the importance of oxygen vacancies themselves in the magneto-optical properties of TiO$_2$, we probe bulk and quasi-spherical nanocrystals of this material by quantum mechanical first-principles computations in the phases of rutile, anatase, and brookite. We show that the impurity electrons related to an oxygen vacancy are trapped by two of the three neighboring Ti$^{+4}$ cations which thereby are reduced into Ti$^{+3}$. These defect states appear in the band gap as polaron states which affect the magnetic and optical properties of reduced TiO$_2$ systems. 

\section{Method}
The localized character of polarons, when treated with standard density functional theory (DFT) methods, is roughly described due to the self interaction error inherent to local and semilocal approximations. This error is expressed as a deviation from the correct Koopmans behavior of the highest electronic level; this level, in defective titania, corresponds to an impurity state\cite{koops1}. To overcome the problem, we resorted to a self-interaction corrected DFT method \cite{koops2,hannes,lany,lany2,lany3,mio3} known as NLEP, which is computationally less demanding than other methods such as those based on hybrid exchange correlation functionals\cite{hybrid} and Green's functions\cite{gw}, and which corrects the band gap problem inherent to local and semi-local DFT approximations. In the NLEP formalism, non-local external potentials and $U$ Hubbard terms are applied to Ti and O atomic orbitals. For computations we used the VASP code\cite{kresse-1,kresse-2,kresse-3,paw}, a plane-wave cutoff energy of 300 eV, an exchange correlation functional defined in the generalized gradient approximation of Perdew, Becke, and Ernzerhof \cite{pbe}, and Ti$_{\rm{pv}}$ and O$_{\rm{s}}$ pseudopotentials accounting for 10 and 6 electrons, respectively. The bulk first Brillouin zone was sampled with a shifted Monkhorst-Pack grid of $4 \times 4 \times 4$  $k$ points, and the quantum dots were placed in supercells with 10 {\AA} of vacuum space and calculated at the gamma point. The atomic structures were relaxed until the forces on the individual nuclei became smaller than 0.02 eV/{\AA}.

\begin{figure}
\includegraphics*[scale=.45]{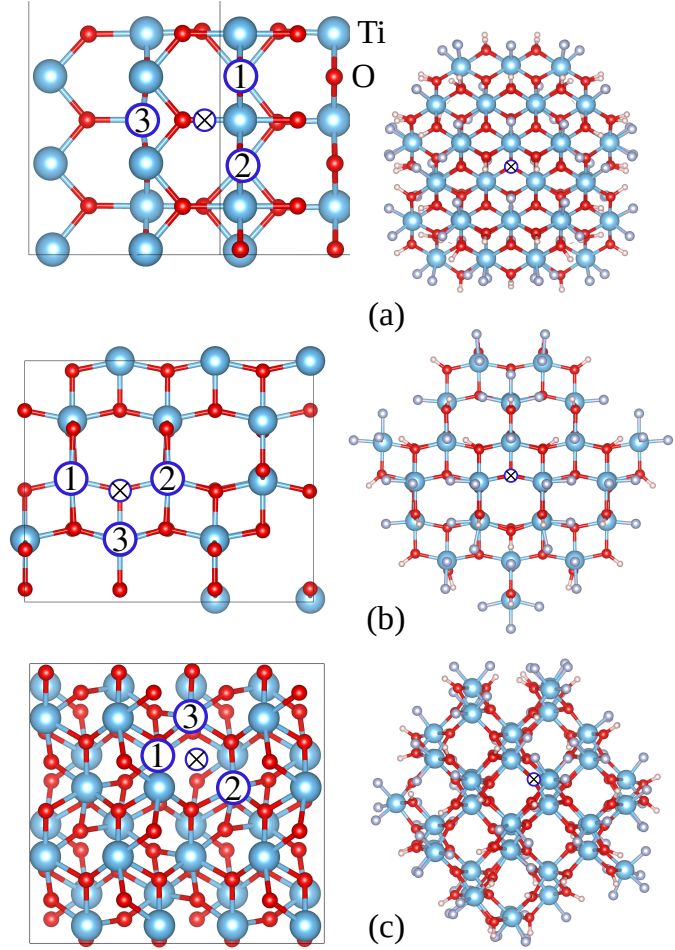}
\caption{\label{fig:structures}Atomistic description of bulk (left) and nanoparticles (right) of reduced TiO$_2$ in (a) rutile, (b) anatase, and (c) brookite phases. The numbers label the three neighboring Ti sites by the vacancy place, which is plotted as a yellow sphere.}
\end{figure}

\section{Description of bulk and nanoparticle systems}
We study rutile, anatase, and brookite crystals of bulk TiO$_2$ composed of 72, 108, and 96 atoms in supercells of $2\times2\times3$, $3\times3\times1$, and $1\times2\times2$ unit cells, respectively. We use optimized lattice parameters which are close to the experimental ones \cite{ar,b}; they are included in the SI. Because we have a single oxygen vacancy per supercell, the defect concentration amounts to $\sim2 \times 10^{21}$ cm$^{-3}$, comparable to the one in highly reduced TiO$_2$ \cite{high}. Furthermore, we calculate rutile, anatase, and brookite quasi-spherical nanocrystals of about 1.6 nm size and about 280 atoms, as shown in Fig. \ref{fig:structures}. To avoid the appearance of surface states in the band gap \cite{jesus}, pseudohydrogens of fractional charges\cite{pass} H$_{4/3}$ and H$_{2/3}$ are attached, respectively, to the edge titanium and oxygen atoms.

\begin{figure}
\includegraphics*[scale=0.35]{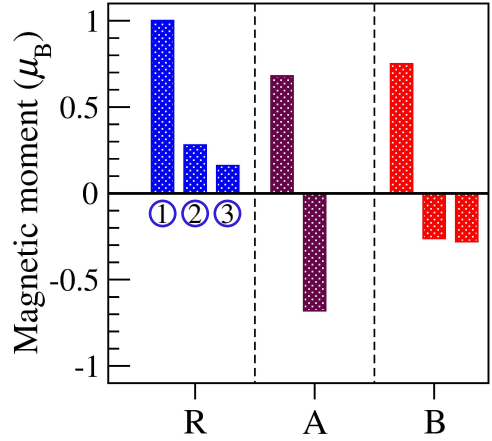}
\caption{\label{fig:local}Local magnetic moments at the three Ti sites surrounding the O vacancy in bulk rutile (R), anatase (A), and brookite (B). The given numbers indicate the positions of the Ti atoms as labeled in Fig. \ref{fig:structures}.}
\end{figure}

\begin{figure}
\includegraphics*[scale=0.5]{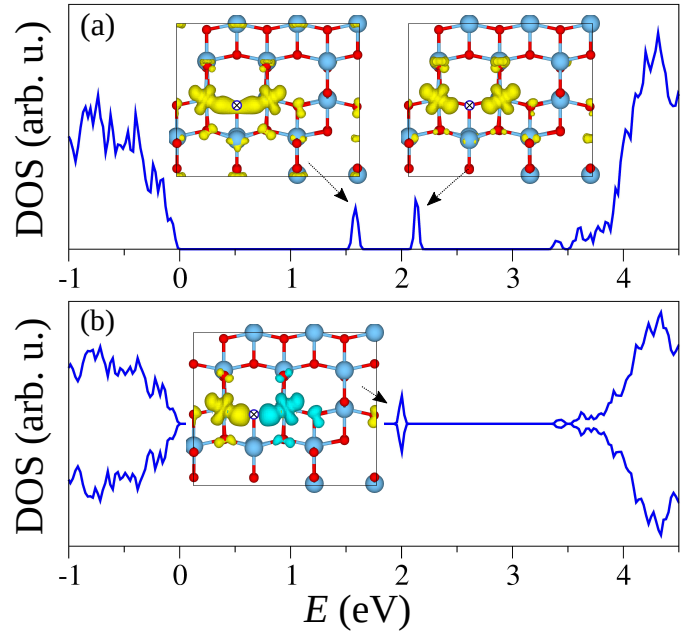}
\caption{\label{fig:dos}Densities of states (DOS) in bulk anatase TiO$_2$ with bipolarons in either (a) spin-triplet or (b) spin-singlet configuration. Insets show isosurfaces of constant spin density (0.0035 $a_0^{-3}$) with light (yellow) color when the spin is up and dark (blue) color when it is down. As before, the oxygen vacancy is represented as a yellow sphere.}
\end{figure}

\section{Results and Discussion}

\subsection{Magnetic properties}
First, we investigate oxygen defects, related polarons, and magnetic coupling properties in the bulk phases. The formation of an oxygen vacancy yields two impurity electrons which create the bipolaron states at the neighboring Ti sites in either spin-singlet or spin-triplet configuration, resulting in a total magnetic moment in the unit cell of zero and two Bohr magneton ($\mu_{\rm{B}}$), respectively. The polaronic nature of these defects emerge as large local magnetic moments and concomitant structural changes at the hosting Ti positions, as reported in Fig. \ref{fig:local}. In the quantum dots, the local magnetic moments are closer to one Bohr magneton due to a loss in the number of lattice atoms around the vacancy. 

\begin{figure}
\includegraphics*[scale=0.35]{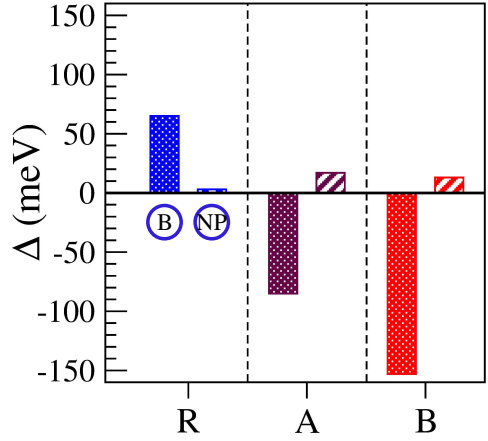}
\caption{\label{fig:delta}Energy difference $\Delta=E^S-E^T$ between bipolaron spin-singlet (S) and spin-triplet (T) configurations for vacancies placed in the bulk (B, solid bars) and at a nanoparticle central site (NP, stripped bars).}
\end{figure}

In Fig. \ref{fig:dos} we plot the densities of states for the anatase phase. It shows the presence of defect states and a 3.3 eV band gap close to the experimental one of 3.2 eV\cite{bg1}. The densities of states in bulk rutile and brookite are supplied in the SI, with band gaps of 3.1 and 3.4 eV in consonance with the experimental gaps of 3.0 and 3.14 eV\cite{bg,bg1}, respectively. In bulk anatase TiO$_2$, the spin-triplet state is formed by two electrons placed at bonding and antibonding states arising from the hybridization of two neighboring Ti-$d$ orbitals (as labeled with numbers 1 and 2 in Fig. 1b). Moreover, the positions of the electronic energy levels in spin-singlet and spin-triplet states are consistent, on one hand, with the strong localized nature of polarons in bulk anatase TiO$_2$ and the consequent larger stability of the spin-singlet configuration, see Fig. \ref{fig:delta}, and, on the other hand, with comparable structural distortions around the vacancy for spin-singlet and triplet states. Previous calculations based on a hybrid density-functional method yielded, however, a spin-triplet ground state \cite{hybrid}. Analogous results on bulk rutile and brookite crystals can be found in the Supplementary Information. In Fig. \ref{fig:delta} we report the energy differences between spin-singlet (S) and spin-triplet (T) configurations, $\Delta=E^S-E^T$. The ground state for bipolarons in bulk rutile titania is a spin-triplet state. Furthermore, since $\Delta$ exceeds the thermal energy at room temperature ($k_{\rm{B}}T\simeq$26 meV), polaron spins in bulk rutile TiO$_2$ may promote room-temperature ferromagnetism, in agreement with preliminary measurements\cite{cristina}. For bulk anatase and brookite crystals, however, the ground state for bipolarons is a spin-singlet state. Hence, the ferromagnetic response observed in anatase TiO$_2$ would have another origin, possibly related with F$^+$ color centers characterized by a single electron at the vacancy site displaying a paramagnetic signal\cite{color}.

As for quantum dots, we investigate oxygen vacancies placed at central sites and below the surface saturated by pseudohydrogens (Fig. \ref{fig:structures}). The computed total energy decreases when the vacancy is moved from the center to the surface by 0.53, 0.59, and 0.87 eV in rutile, anatase, and brookite nanocrystals, respectively, in agreement with previous theoretical results which show an energy decrease of about 0.5 eV.\cite{mio2} In addition, $\Delta$ is almost null for vacancies by the nanoparticle surface. From Fig.\ref{fig:delta} we conclude, interestingly, that independently of the crystal phase the ground-state spin configuration for bipolarons in quantum dots is a spin-triplet state. As a consequence, oxygen vacancies in these small systems would produce magnetism, and this effect appears to be just induced by the crystal size.

\subsection{Optical properties}
\begin{figure}
\includegraphics*[scale=.75]{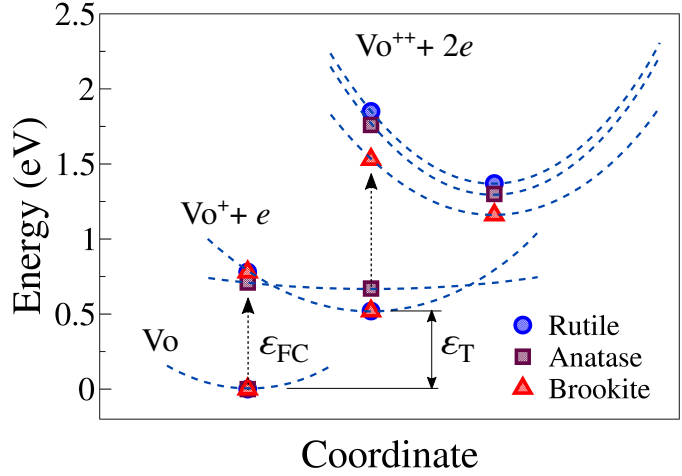}
\caption{\label{fig:coordinate}Configuration coordinate diagram \cite{valle} illustrating the optical ($\varepsilon_{\rm{FC}}$) and thermodynamic ($\varepsilon_{\rm{T}}$) transitions of impurity electrons to the conduction band.}
\end{figure}

Second, we investigate the optical properties of reduced titania by addressing the transition energies of defect electrons to the conduction band. We focus on the transitions V$_{\rm{O}}\rightarrow\rm{V}_{\rm{O}}^+$ and V$_{\rm{O}}^+\rightarrow\rm{V}_{\rm{O}}^{++}$ (Fig. \ref{fig:coordinate}). In the Franck-Condon approximation, these optical levels are expressed as

\begin{equation}
 \varepsilon_{\rm{FC}}\rm{(+1/0)}= E(V_{\rm{O}}^+:V_{\rm{O}})-E(V_{\rm{O}})
\end{equation}

and

\begin{equation}
 \varepsilon_{\rm{FC}}\rm{(+2/+1)}= E(V_{\rm{O}}^{++}:V_{\rm{O}}^{+})-E(V_{\rm{O}}^+),
\end{equation}

\noindent where $E$(V$_{\rm{O}}^+$:V$_{\rm{O}}$) is the total energy of a defective crystal with a neutral vacancy V$_{\rm{O}}$, frozen atomic structure, and one electron in the conduction band, and $E$(V$_{\rm{O}}^{++}$:V$_{\rm{O}}^{+}$) is the total energy of a defective crystal with an excited vacancy V$_{\rm{O}}^{+}$, frozen atomic structure, and two electrons in the conduction band, see Fig. \ref{fig:coordinate}. The thermodynamic transitions levels, however, involve fully relaxed structures and for this reason they are shallower than the optical ones; we express them as

\begin{equation}
 \varepsilon_{\rm{T}}\rm{(+1/0)}= E(V_{\rm{O}}^+)-E(V_{\rm{O}})
\end{equation}

and

\begin{equation}
 \varepsilon_{\rm{T}}\rm{(+2/+1)}= E(V_{\rm{O}}^{++})-E(V_{\rm{O}}^+),
\end{equation}

\noindent For bulk rutile and brookite, the optical levels lie significantly deeper in the band gap than the thermodynamic ones. However, for anatase, both optical and thermodynamic levels are close to each other due to a small excited-state relaxation of the lattice (Fig. \ref{fig:coordinate}). Interestingly, for bulk rutile, $\varepsilon_{\rm{FC}}$(+1/0)=0.78 eV and $\varepsilon_{\rm{FC}}$(+2/+1)=1.33 eV are in close agreement, respectively, with the 0.75 eV and 1.18 eV absorption peaks reported in the experiments.\cite{nanoscale,photo,opticrut1,opticrut2,spectr} We note that $\varepsilon_{\rm{FC}}$(+1/0)=0.78 eV is significantly closer to the experimental level than the previous value of 0.47 eV calculated with a hybrid density-functional method\cite{hybrid}. For bulk anatase, $\varepsilon_{\rm{FC}}$(+2/+1)=1.09 eV coincides with the 1.1 eV peak obtained from resonant photoemission and x-ray absorption spectroscopy\cite{nanoscale,photo,spectr}. As for the quantum dots, we study the transitions V$_{\rm{O}}\rightarrow\rm{V}_{\rm{O}}^+$ and calculate $\varepsilon_{\rm{FC}}$(+1/0), which is 1.76, 1.71, and 1.30 eV for rutile, anatase, and brookite phases, respectively. These values exceed the bulk ones because they are enhanced by the confinement of carriers within these quantum nanosystems.

\section{Conclusions}
We investigated the formation of polarons in reduced titania due to oxygen vacancies and their relevance to the magneto-optical properties such as the absorption spectrum. Based on a self-interaction corrected density functional theory method, we addressed both bulk and nanoparticles of TiO$_2$ in rutile, anatase, and brookite phases. Interestingly, the ground state for bipolarons in bulk rutile TiO$_2$ as well as in quantum dots is a spin-triplet state, what suggests that oxygen vacancies may yield a magnetic signal out of this oxide compound. Furthermore, we reported optical and thermodynamical transition levels from defect states in the band gap to the conduction band. For bulk rutile titania, the first optical transition level is at 0.78 eV, in good agreement with the 0.75 eV experimental value\cite{opticrut1}. This result confirms the suitability of our theoretical approach for the study of optoelectronic properties in oxide compounds.

\bibliography{bibliography}
\end{document}